\documentclass[12pt]{article}
\usepackage{a4}
\newcommand{\be}{\begin{equation}}
\newcommand{\ee}{\end{equation}}
\newcommand{\bea}{\begin{eqnarray}}
\newcommand{\eea}{\end{eqnarray}}

\begin{document}

\begin{center}
\title{Baryogenesis and Polyakov line condensate during inflation}
\author{As. Abada$^a$, M.B. Gavela$^b$ and O. P\`ene$^a$} \par
\maketitle
{$^a$ Laboratoire de Physique Th\'eorique et Hautes
Energies\footnote{Laboratoire
associ\'e au
Centre National de la Recherche Scientifique - URA D00063
\\e-mail: abada@qcd.th.u-psud.fr, gavela@delta.ft.uam.es,
pene@qcd.th.u-psud.fr.}}\\
{Universit\'e de Paris XI, B\^atiment 211, 91405 Orsay Cedex,
France}\\$^b$Departamento de F\' \i sica Te\'orica, Universidad Aut\'onoma
de Madrid, Canto Blanco, 28049 Madrid.
\end{center}

\begin{abstract}

The thermal average of the Polyakov line is non-zero at high temperature. It
has been suggested that
 its phase could be relevant for baryogenesis within the standard model,
providing a source of
spontaneously broken CP violation.
This scenario suffers from a causal domain problem which cannot be cured by
 inflation: quantum fluctuations in de Sitter space overwhelm the periodicity
 of the potential.

\end{abstract}
\begin{flushright} LPTHE Orsay-96/11\\ FTUAM FEV/96/6\\ hep-ph/9602388
\end{flushright}
\newpage

\section{Introduction}

The origin of the absence of primordial antimatter in the universe (or at least
in the Virgo cluster of galaxies) remains unclear. The observed baryon to
photon density ratio, $n_B/n_\gamma\sim 10^{-9}$, although small, is difficult
to explain.

The most promising avenue to address this problem is baryogenesis: starting
from a matter-antimatter symmetric universe, the microscopic laws of physics
should
affect the cosmological evolution so as to result in the above mentioned
experimental observation.
 For this to happen \cite{sakharov}, the presence of C, CP and baryon number
(B) violating processes
during an out-of-equilibrium phase of the expansion is required.

 A baryogenesis mechanism based exclusively on the Standard Model is
most appealing. It has been argued \cite{kuzmin} that an out-of-equilibrium
mechanism would be provided by a first order electroweak phase transition,
although the pertinent dynamical conditions seem to be excluded by the present
bound on the Higgs mass \cite{jansen}.
Assuming such a transition,  B-violating transitions stem from the anomaly
\cite{thooft}, \cite{kuzmin}. The remaining question is whether the Standard
Model  provides with sufficient strength the rest of the ingredients, that is,
C and CP violation, the latter being the most difficult issue.

 Two well-known sources of CP violation are potentially present in the standard
model, both explicit in character. The QCD $\theta$ parameter, if different
from zero, is one of them, albeit a C conserving one. No mechanism based on it
has been sucessfully proposed. The other one is of electroweak origin, the
phase in the Cabibbo-Kobayashi-Maskawa matrix. This road has been attempted
\cite{shapo}, although  it has been shown \cite{belen} that the thermal noise
in the quark-gluon plasma reduces the resulting baryon asymmetry in this
scenario to $n_B/n_\gamma\le 10^{-20}$, less than ten orders of magnitude below
the needed figure.

It has been recently realized that two additional sources of CP violation are
present in the Standard Model at
 {\it finite temperature}, due to the related breaking of Lorentz invariance.
Both are examples
of spontaneous breaking of CP violation and will then have to face the issue of
uncorrelated CP phases
 in causally disconnected domains, that is  whether the typical size of such
domains is larger than the present
 horizon or, at least, larger than a typical cluster. One of these scenarios is
based on $Z$ meson condensates
 \cite{turok}, we will not dwell on it here.

The
 other one \cite{dixit}-\cite{korthals} is based on thermal effects of the
phase of the Polyakov line:
\bea
\mathrm{P}(\vec x)=\frac 1 n \mathrm{Tr}\bigr\{ {\cal P}e^{ig\int_0^\beta
A_0(\vec x,t)dt} \bigl \},
\label{ligne}
\eea
where $A_\mu$ is the $\mathrm{SU(N)}$
gauge field, $g$ is the coupling constant, ${\cal P}$ denotes the path ordering
and
$\beta=1/\mathrm{T}$, $\mathrm{T}$ being the temperature.

 The thermal average of the Polyakov line, $<\mathrm{P}(\vec x)>_\mathrm{T}$,
is  a gauge invariant
quantity and constitutes an order parameter for confinement  in pure Yang-Mills
theory \cite{polyakov}, i.e. {\it when only gauge fields
 are present}. The action presents the well-known $\mathrm{Z}(N)$ discrete
symmetry whose  effect is to multiply the
average \break $<\mathrm{P}(\vec x)>_\mathrm{T}$ by one of the numbers:
$e^{{-2\pi i k \over N}} \, , \ k=0,1,...,N-1$\footnote{The group constituted
by the elements:
$e^{{-2\pi i k \over N}}\mathbf{1} \, , \ k=0,1,...,N-1$, is the center of $
\mathrm{SU(N)}$.}. At low
 temperature $<P(\vec x)>_\mathrm{T}= 0$, and the theory confines. At high
temperature, the symmetry is
 spontaneously broken, $<\mathrm{P}(\vec x)>_\mathrm{T}\ne 0$, corresponding to
the nonconfining phase.

At high temperature it is appropiate to define an effective potential for the
quantity $<\mathrm{P}(\vec x)>_\mathrm{T}$ \cite{weiss}, which presents $N$
degenerate minima: $<P(\vec x)>_\mathrm{T}=e^{{-2\pi i k \over n}}\, ,
k=0,1,...,N-1$, reflecting the discrete $\mathrm{Z}(N)$ symmetry. Those minima
with non-trivial phases are putative sources of CP violation.

 In the presence of fermions the degeneracy is lifted \cite{dixit}, and
 $<\mathrm{P}(\vec x)>_\mathrm{T}= 1\,\ (k=0)$ in the true vacuum. The
remaining local minima are metastable. In its cosmological evolution, the
universe could have fallen in one of those at some GUT or Planck mass scale
\cite{ignatius}, \cite{korthals}.
 In the standard cosmology these metastable minima decay by tunneling
\cite{coleman}.
The authors of \cite{bhatta} argue that they are long-lived and
have not decayed by the time of the electroweak transition i.e. $10^2\,
\mathrm{GeV}$.

There are some serious objections \cite{belyaev}-\cite{smilga} to this
hypothesis.
The domains corresponding to the minima found in the potential of  $<P(\vec
x)>_\mathrm{T}$, or the barriers (walls)  separating them, might not be
interpreted as real physical objects in Minkowski space. This issue
deserves a careful study not attempted in this work, although it will
eventually surface  disguised in
 the form of a ``complex chemical potential".

In this paper we will address the issue of causality.
The ``choice" between the metastable or stable vacua being spontaneous, one
same vacuum
cannot cover a region larger than a causally connected domain at the time
of  spontaneous symmetry breaking, i.e., necessarily before the electroweak
phase transition.
In a standard expansion of the universe, not inflationnary, such a causally
connected domain evolves
 eventually into a region whose size today does not exceed a few million of
kilometers,
 much smaller than the size of a cluster of galaxies, $\simeq 25$ Mpc.
Such a spontaneously broken CP symmetry could not explain that the baryon
number asymmetry
 keeps the same sign at least all through our cluster of galaxies.

There are two possible ways out:
\begin{enumerate}
\item A mechanism of explicit breaking of the CP symmetry which tilts the
spontaneous one, as for instance advocated in \cite{turok} for the $Z$
condensate. We know of no such mechanism in the scenario under discussion.
\item Inflation.
If the domains have formed  long enough before the end of an inflationary
period, one might
expect them to grow into a region
much larger today than a cluster of galaxies\footnote{During inflation, a small
and causally connected patch of size
 $\le H^{-1}$  grows exponentially to encompass
 the entire today's universe.}. The purpose of this paper is to decide whether
this might have happened or not.
\end{enumerate}

 A priori, as the temperature falls exponentially
 during inflation,  it can be expected that the effective potential for the
condensate flattens. Thermal
 fluctuations should become negligible as well. Quantum fluctuations in de
Sitter space may allow the
 growth of the condensate, which could then get arbitrarily large values by the
end of inflation.
Upon reheating, the effective potential starts to wiggle again,  and the
condensate
could fall into the nearest minimum\footnote{ This is in fact a delicate point,
as the dynamics of
 reheating is a difficult problem which remains to be explored.
 The conclusions of our paper
make this issue
irrelevant.}.

 The question to decide is whether the ensemble of then causally disconnected
domains which forms our universe
 today (or our cluster) has fallen into the same minimum, or else have a random
distribution, incompatible
 with observation. Those domains, disconnected at the end of inflation, have
certainly belonged to a common
 causally connected one, some time in the past, during inflation. The issue
is whether this common memory forces them to fall into the same minimum of the
effective potential during reheating.

This type of problem has been extensively studied for scalar field potentials.
The new elements to take into account are:

\noindent 1) The fact that the condensate, although described by an effective
potential, is not alike to a simple dynamical field, such as the inflaton.

\noindent 2) The periodicity in the phase of the effective potential, and the
dependence of the period on temperature, which will play a major role.

 The scenario \cite{korthals}  uses the imaginary time thermofield formalism
\cite{bernard} in
which the real time is complexified:
$t\to i\tau$. The ``evolution" in the imaginary time direction describes the
Boltzman
distribution $e^{-\beta H}$ ($\beta = 1/T$), and the real time disappears since
one describes
a stationnary thermal equilibrium.

During inflation, in order to consider the interplay of thermal effects with
quantum ones, it
 is necessary to
 develope some tools for treating gauge fields at finite temperature in de
Sitter space. Even
 when the
 temperature becomes negligible, the thermal  density matrix is relevant to
understand the
beginning and the
 end of inflation. More important, a redshifted density matrix remains during
inflation as the
 relic from the
 pre-inflationary  thermal one. A real time variable has to be kept as a record
of the
 cosmological time. The choice of a complex time variable, for temperature
purposes, is non
 trivial as many possible choices are possible in cosmology \cite{peeble},
which would lead to
 very different results after complexification. We will have to find our way
through these
 meanders.

We solve this puzzle starting from basics. We derive the thermal density matrix
that
 generalizes the Boltzman distribution. We then investigate whether any time
variable has an
 evolution operator that allows to describe the latter density matrix via
complexification,
 and we determine it.

The complete scenario requires to consider the effective potential for
$SU(3)\times
 SU(2)\times U(1)$ with fermions. Our physical results are general enough so as
to be present
 in a simpler field theory. For this reason we will perform the computations
for
a  pure gauge $SU(2)$ Lagrangian. All throughout the paper, the gravitational
effects of the
gauge fields are neglected, that is, the vacuum energy is approximated by the
inflaton
contribution and taken as constant.

In section \ref{fock}, we derive the field equations for the gauge fields in
de Sitter space, and
build the corresponding Fock space. In section \ref{potential}, the free energy
both with real and imaginary
time is computed, and an effective potential with imaginary time follows.
In section \ref{instability}, we study the stability of the metastable vacua
for thermal and quantum fluctuations and finally we conclude.

\section{Fock expansion of the gauge fields}\label{fock}

\subsection{Metric}\label{metric}
During inflation, the scale factor $a(t)$ grows exponentially and the metric
can be written in the form:
\bea
ds^2= g_{\mu\nu}dx^\mu dx^\nu= dt^2- e^{2Ht} (dx^2+dy^2+dz^2)\label{ds}.
\eea
where $t$ is the proper time measured in the comoving frame where the spatial
variables are constant, and therefore denoted comoving variables, which means
that a massive object at rest at the position $\vec x$ will keep the coordinate
$\vec x$ in later times. $H$ is the expansion rate ${\dot a(t)/ a(t)}$ of the
universe, that is, the Hubble``constant'', and the metric tensor $ g_{\mu\nu}$
is given by
\begin{equation}g_{\mu\nu}=  \left( \begin{array}{cccc}
1&0&0&0\\0&-e^{2Ht}&0&0\\0&0&-e^{2Ht}&0\\0&0&0&-e^{2Ht}\\
\end{array}\right)\ \ ; \qquad g^{\mu\nu}=  \left( \begin{array}{cccc}
1&0&0&0\\0&-e^{-2Ht}&0&0\\0&0&-e^{-2Ht}&0\\0&0&0&-e^{-2Ht}\\
\end{array}\right)\label{metrique}.
\end{equation}
Its determinant is
\bea{
det\{g_{\mu\nu}\}=\sqrt{-g}= e^{3Ht}.
}\eea

It is convenient to express the metric in terms of the ``cosmic time" defined
by
$d\eta = {dt/ a(t)}+b$, where the scale factor has the form $a(t)=a_0e^{Ht}$.
 The origin for $t$ and $\eta$ is chosen so that $b=0$ and $H\eta = -1$ at
the beginning of inflation.
The cosmic time,
\bea
\eta=-H^{-1}e^{-Ht},
\label{conformal}\eea
is such that the metric takes the form
\bea ds^2=a(\eta)^2(d\eta^2-(dx^2+dy^2+dz^2)),\label{cosmic}\eea
which explains why  $\eta$ is often dubbed ``conformal time".

It is sometimes useful to use the so-called physical variables, where the
proper time is kept, and $d\vec x_{phys}=e^{Ht} d\vec x$. All computations will
be performed in the comoving variables, the others will be referred to in
physical discussions.

\subsection{Field equations}\label{field}

We assume that the inflationary process is driven by a scalar ``inflaton"
field.
Within this space, it is possible to study the field equations for gauge and
fermion fields, and it is consistent to simultaneously neglect the effect of
those fields on the inflation rate. The latter is dominated by  the ``inflaton"
ones and  will be taken as constant during inflation\footnote{ Indeed, it will
be shown that the gauge field contribution to the energy-momentum tensor
vanishes as $e^{-4Ht}$.}.

In flat space, the pure $SU(2)$ Lagrangian density is
\bea
{\cal L}= -\frac 1 4 G_{\mu\nu}^aG^{a\mu\nu }\label{lag},
\eea
where $a =1,2,3$ labels the isospin. In this paper we shall restrict ourselves
to the $SU(2)$ gauge group, although a generalization to $SU(N)$ is
straightforward.

In de Sitter space it is convenient to define as well the Lagrangian
\bea
{
{\cal L} ^{\mathrm{DS}}=\sqrt{-g} {\cal L}=-{1\over 4} e^{3Ht} G_{\mu
\sigma}G_\nu^\sigma g^{\nu\mu},
\label{lagrangien}
}
\eea
so that the action is given by
\bea S=\int d^4x\, {\cal L} ^{\mathrm{DS}}.\eea
${\cal L} ^{\mathrm{DS}}$ can be expressed in a non-covariant way,
\bea
{\cal L} ^{\mathrm{DS}}={\sqrt{-g}\over 2} \sum_{a=1}^{3} \bigr [
e^{-2Ht}\sum_{l=1}^{3}  {G_{0\ell}^a}^2- e^{-4Ht}\sum_{j<k}^3 {G_{jk}^a}^2
\bigl  ].
\label{lagnoncov}
\eea

Developing eq. (\ref{lagnoncov}) up to terms quadratic in the fields,
\bea
{\cal L} ^{\mathrm{DS}}={\sqrt{-g}\over 2}
\biggr \{
e^{-2Ht} \sum_{i=1}^{3}
&\biggr \{&
\sum_{a=1}^{3}
\bigr \{
\bigr(\partial_0 A_i^a -\partial_i A_0^a +(A_0 \wedge A_i)^a\bigl  )^2
\bigl\}
\biggl\}\cr
-{1\over 2}e^{-4Ht} \sum_{j,k}^3 &\biggr\{& \sum_{a=1}^{3}(\partial_j
A_k^a-\partial_k A_j^a)^2\biggl\}
+ ....\biggl \}.
\label{lagexplicite}
\eea

Higher order terms are neglected as, at high temperature, the interaction is
assumed to be weak.
We will work in the temporal gauge $A_0^a=0, a =1,2,3$.

 In terms of the conformal time, eq. (\ref{conformal}), the de Sitter
Lagrangian writes:
 \bea
{\cal L} ^{\mathrm{DS}}={e^{-Ht}\over 2}
\biggr \{
 \sum_{i=1}^{3}
&\biggr \{&
\sum_{a=1}^{3}
\bigr \{
\bigr(\partial_\eta A_i^a\bigl )^2
\bigl\}
\biggl\} \cr
-{1\over 2} \sum_{j,k}^3 &\biggr \{& \sum_{a=1}^{3}
(\partial_j A_k^a-\partial_k A_j^a)^2\biggl\}
+ ....\biggl \}
\label{lageta}.
\eea
It is noticeable that this Lagrangian  is identical to the
flat space one up to the global factor $e^{-Ht}$.
For later use we have chosen a privileged direction in isospin space, $\hat 3$,
and the following convention for the transverse gauge fields:
\bea
A_i^{\pm}= {1\over \sqrt{2}}(A_i^1{\pm}i A_i^2).
\eea

The equations of motion for the fields, $A_i^3$ and $A_i^{\pm}$ can be obtained
either
 by varying the action
($S=\int d^4x \sqrt{-g}{\cal L} $), or using the conservation of the
energy-momentum tensor, $T^{\mu\nu}$$_{;\nu}$.

 For $A_i^{a}$ ($a=+,-,3$) the evolution equation is\footnote{It is necessary
to pay careful attention to the position of the spatial index. Because of the
form of $g^{\mu\nu}$, eq. (\ref {metrique}), $\dot f_i$ corresponds to ${d\over
dt}f_i$ only when the index $i$ is down, and $\overrightarrow{\nabla} \cdot
\overrightarrow{A^{a}}=\sum_{j}\partial_j A_j^{a}$.}

\bea
{\ddot A}_i^{a} +H{\dot A}_i^{a}  -
e^{-2Ht}\bigr( \Delta A_i^{a}-\partial_i (\overrightarrow{\nabla} \cdot
\overrightarrow{A^{a}}) \bigl )=0.
\label{equaAig}
\eea

 The transversally polarized components of $A_i^{a}$,  verify
\bea
{\ddot A}_i^a +H{\dot A}_i^a  - e^{-2Ht}
\Delta A_i^a=0,
\label{equaAi3t}
\eea
and we will restrict ourselves to their study from now on, since only radiation
modes are relevant in this problem. In terms of the conformal time, eq.
(\ref{equaAi3t}) reduces to

\bea
{\partial ^2 A_i^a \over \partial \eta^2} +\Delta A_i^a =0. \label{equaAi3eta}
\eea

Before expanding the gauge fields in momentum space, it is necessary to
determine
the conjugate fields in order to fix the normalization of the solutions.
\bea
\pi^{\mu(a)}={\partial {\cal L}^{\mathrm{DS}}\over\partial (\partial_0 A_\mu^a)
};
\cases{\pi^{0(a)}=0\cr \pi^i=E^{i(a)}}
\label{lepi}\eea
In analogy with $A_i^1$ and $A_i^2$, the transverse conjugate fields, $E_i^1$
and $E_i^2$, are defined by
\bea
E_i^{\pm}= {1\over \sqrt{2}}(E_i^1{\pm}i E_i^2),
\eea
and the corresponding conjugate fields, $E_i^{a}$, are
\bea
E_i^{a}=-e^{3Ht}\bigr(\dot A_i^{a} \bigl )=-\frac 1 {\eta^2 H^2}\frac
\partial{\partial \eta}A_i^{a}. \label{Ei}
\eea
 The standard canonical commutation relations are still valid in de Sitter
space,
\bea
\bigr [ A_i^{a}, E_j^{b}\bigl ]=i g_{ij}\delta^3(\vec x-\vec y)\delta_{ab}=-i
\delta_{ij}\frac 1 {\eta^2 H^2}\delta^3(\vec x-\vec
y)\delta_{ab}.\label{commutateur}
\eea

\subsection{Gauge field expansion into creator and anihilators}\label{gauge}

 It will be shown that the solutions of the field equations are analogous
 to those in flat space, with the replacement of $t$ by the cosmological time
$\eta$.

\bea
A_i^{3}(\vec x, t)=\int d^3 k \sum_{\lambda=1,2}\bigr [
{\cal A}_k^{\lambda 3}(\eta) \varepsilon_{i}^{\vec k,\lambda} {\mathbf{
a}}^3(\vec k,\lambda)e^{i\vec k \cdot \vec x}
+
{\cal A^*}_k^{\lambda 3}(\eta) \varepsilon_{i}^{\vec k,\lambda}
{\mathrm{\mathbf{ a}}}^{3\dagger}(\vec k,\lambda)e^{-i\vec k \cdot \vec
x} \bigl ]
\label{devhenkela}
\eea

\bea
A_i^{\pm}(\vec x, t)=\int d^3 k \sum_{\lambda=1,2}\bigr [
{\cal A}_k^{\lambda \pm}(\eta) \varepsilon_{i}^{\vec k,\lambda}
{\mathrm{\mathbf{ a}}^\pm}(\vec k,\lambda)e^{i\vec k \cdot \vec x}
+
{\cal A^*}_k^{\lambda \mp}(\eta)\varepsilon_{i}^{\vec k,\lambda}
{\mathrm{\mathbf{ a}}^{\mp\,\mathbf{\dagger}}}(\vec k,\lambda)e^{-i\vec k
\cdot \vec x} \bigl ]
\label{devhenkelpm}
\eea
 In eqs. (\ref{devhenkela}) and (\ref{devhenkelpm}), the  $\sum$  extends upon
the transversal
polarization modes, with the corresponding vectors,
$\varepsilon_{i}^{\vec k,\lambda}$, chosen to be real and normed,
$\varepsilon_{i}^{\vec k,\lambda}
\varepsilon_{j}^{\vec k,\lambda}=\delta_{ij}$, for $\lambda=1,2$.

 The annihilation and creation operators satisfy the commutation relations
\bea \cases{
\begin{array}{l}
\bigr [
\mathbf{a}^b(\vec k,\lambda),\mathbf{a}^{c\,\mathbf{\dagger}}
(\vec k',\lambda')\bigl ]
=\delta_{bc} \delta_{\lambda\lambda'}\delta^3(\vec k-\vec k') \cr
\bigr [
\mathbf{a}^{b\,\mathbf{\dagger}}(\vec k,\lambda),
\mathbf{a}^{c\,\mathbf{\dagger}}(\vec k',\lambda')
\bigl ]=
\bigr [
\mathbf{a}^b(\vec k,\lambda),
\mathbf{a}^c(\vec k',\lambda')
\bigl ] =0.
\end{array}}
\label{commutation}
\eea

{}From eq. (\ref{equaAi3eta}),
\bea
{d^2\over d\eta^2}{\cal A}_k^{\lambda a}(\eta)  +k^2{\cal
A}_k^{\lambda a}(\eta)=0,
\label{equaAk3}\eea
with solution
\bea
{\cal A}_k^{\lambda a}(\eta)=a e^{ik\eta} +b e^{-ik\eta},\eea
where $k=|\vec k|$. For $Ht\to 0$, the matching with the flat space solution,
$\propto e^{-ikt}$,  requires $a=0$. Imposing finally
the canonical
commutation relations, eq. (\ref{commutateur}), it results
\bea
{\cal A}_k^{\lambda a}(\eta)={1\over (2\pi)^{3/2}}{1\over \sqrt{2k}}
e^{-ik\eta}.
\label{henkelA3}\eea

\section{The effective potential for the Polyakov line}\label{potential}

Our goal in this section is to determine the thermodynamical potential
corresponding to the gauge fields, in order to derive the effective potential
for the Polyakov line,
\bea
\mathrm{V_{eff}(<P(\vec x)>)}=\frac {F(\mu,V,T)}{V}. \label{signe} \eea
For this, it suffices to compute the energy-momentum tensor. We recall that
the ``thermodynamical potential" of the Grand-Canonical ensemble is given by
\bea
\mathrm{F(\mu, V,T)}= -\beta^{-1}\log{\mathrm{Z_G}},\label{pothermC}
\eea
where $\mathrm{Z_{G}}$ is the partition function of the ensemble, $\mu$ is the
chemical potential, V the volume and T the temperature. From now on the
parameters will be implicit, and $\mu=0$ unless otherwise stated. Recalling
that
\bea
\mathrm{dF}=-\mathrm{N}{d\mu}-\mathrm{p} \mathrm{dV}-\mathrm{S}\mathrm{dT},
\label{differentielleC}
\eea
where $\mathrm{N},\  \mathrm{{p}}, \  \mathrm{S}$ denote  the average number of
modes,
 the pressure and the entropy respectively, it results
 \bea\mathrm{p=-{\partial F\over
\partial V}}, \eea
and in the thermodynamical limit, N $\to \infty$,V $\to \infty$, the function F
becomes the free energy.

The pressure may be deduced from the energy momentum tensor through the
relation
\bea
<\mathrm{
T^\mu_\nu>=diag(\rho,\mathrm{-p},\mathrm{-p},\mathrm{-p})},\label{iso}
\eea
$\rho$ being the energy density and $<\mathrm{T^\mu_\nu}>$ the thermal average
of $\mathrm{T^\mu_\nu}$. In writing (\ref{iso}), the isotropy in the comoving
frames, in which the fluid is instantaneously at rest, has been used.

Furthermore, from the equation of state for radiation\footnote{We shall derive
this equation in de Sitter space in the next subsection.},
\bea
\mathrm{p}={1\over 3}\rho,\label{tiersderho}
\eea
it results for the free energy
\bea
\mathrm{F=-{1\over 3}\rho V},
\label{FCasm}
\eea
and we just need to compute $\rho=<\mathrm{T}_{00}>$. In the following
subsection we will then compute the energy-momentum tensor.

\subsection{Gauge field energy-momentum tensor and the effective
potential}\label{tensor}

The energy-momentum tensor is defined by:
\bea
T_{\mu\nu}={2\over \sqrt{-g}}{\partial (\sqrt{-g}{\cal L})\over\partial
g^{\mu\nu} }-\partial_\rho {\partial (\sqrt{-g}{\cal L})\over \partial
(\partial_\rho g^{\mu\nu})}.
\label{deftmunu}
\eea

\noindent Using
\bea
\frac {\partial \sqrt{-g}}{\partial g^{\mu\nu}}=-\frac 1 2 \sqrt{-g} \
g_{\mu\nu},
\eea
we get
\bea
\mathrm{T_{\mu\nu}}=-G_{\mu\alpha}G_{\nu\beta}g^{\alpha\beta}-g_{\mu\nu}{\cal
L},
\label{valtmunu}
\eea
where the sum over color or weak isospin indices is understood.
 More explicitly:

\bea
\mathrm{T_{00}}&=&-{1\over 2} \bigr [\sum_{i} {G_{oi}^2 g^{ii}} -
\sum_{i<k}{g^{ii} g^{kk} G_{ik}^2}  \bigl ],\\
\mathrm {T_{ii}}&=&- {G_{0i}^2} + \frac 1 2 g_{ii}\left(\sum_l g^{ll}
G_{0l}^2\right)
-\sum_k G_{ik}^2 g^{kk} +\frac 1 2 g_{ii} \sum_{j<k} G_{jk}^2 g^{kk}g^{jj},\\
\sum_{i}\mathrm{T_{ii}}&=&{1\over 2}\bigr [\sum_{i} {G_{oi}^2} -
\sum_{i<k}{g^{kk} G_{ik}^2 } \bigl ].
\eea

 From eq. (\ref{iso}),
\bea
\mathrm{p}=-\sum_{i}{1\over 3 g_{ii}}<\mathrm{T_{ii}}>
\eea
and
\bea
-\sum_{i}{1\over 3 g_{ii}}\mathrm{T_{ii}} = {1\over 3 } \mathrm {T_{00}},
\eea
 verifying the equation of state, eq. (\ref{tiersderho}).

 In order to compute $\rho=<\mathrm{T}_{00}>$, we first expand
$\mathrm{T_{00}}$ over
 the fields $A_i^a$:
\bea
\mathrm{T_{00}}=-{1\over 2} \sum_{i} g^{ii}\biggr\{
\bigr [{\dot A}_i^{+}{\dot A}_i^{-}
+{\dot A}_i^{-}{\dot A}_i^{+}  +({\dot A}_i^{3})^2 \bigl ]\cr
-\frac 1 2\sum_{k} g^{kk} \bigr [(\partial_i { A}_k^{+}-\partial_k { A}_i^{+}
)(\partial_i{ A}_k^{-}-\partial_k{ A}_i^{-})\cr
+(\partial_i {A}_k^{-}-\partial_k{A}_i^{-} )(\partial_i
{A}_k^{+}-\partial_k{A}_i^{+} )
+(\partial_i {A}_k^{3}-\partial_k{A}_i^{3} )^2\bigl ]  \biggl \}.
\eea

As interaction terms have been neglected, the density matrix is diagonal
in the Fock basis. The thermal average of creation and annihilation operators
is then
  $\bigl
<\mathbf{a}^{b\,\mathbf{\dagger}}\mathbf{a}^{b\,\mathbf{\dagger}}\bigr
>=\bigl <\mathbf{a}^b\mathbf{a}^b\bigr >=0 $. $\bigl
<\mathrm{T_{00}}\bigr >$ is thus expressed in terms of $\bigl
<\mathbf{a}^{b\,\mathbf{\dagger}}\mathbf{a}^b\bigr >$ and $\bigl
<\mathbf{a}^b\mathbf{a}^{b\,\mathbf{\dagger}}\bigr >$.

Denoting by $\mathrm{T}_0=1/\beta_0$  the temperature
at the beginning of the de Sitter phase, it results
\bea
\cases{\bigl <\mathbf{a}^b(\vec
k,\lambda)\mathbf{a}^{b\,\mathbf{\dagger}}(\vec k,\lambda)\bigr
>=1+n_k^{b \lambda}(\beta_0)\cr
\bigl <\mathbf{a}^{b\,\mathbf{\dagger}}(\vec
k,\lambda)\mathbf{a}^b(\vec k,\lambda)\bigr >=n_k^{b\lambda}(\beta_0)
},\label{moythermo}
\eea
where $n_k^{b\lambda}$ are the ``Bose-Einstein" distribution functions (we are
in the
``perfect gaz approximation"):
\bea
n_k^{b\lambda}(\beta_0)={1\over e^{\beta_0 k}-1}.
\label{densites}
\eea
In terms of these distribution functions, $\bigl <\mathrm{T_{00}}\bigr >$
writes:
\bea
\bigl <\mathrm{T_{00}}\bigr >= {e^{-4Ht}\over  2\pi^2}\int_0^{\infty}\,dk\,
k^3\sum_{b,\lambda}\left[n_k^{b\lambda}(\beta_0)+\frac 1 2\right],
\eea
where the zero point energy of the vacuum is included, which gives an
ultraviolet divergence.
We will omit them until Sec. 4.
We are left with
\bea
\bigl <\mathrm{T_{00}}\bigr >= 3 {e^{-4Ht}\over  \pi^2}\int_0^{\infty}\,dk\,
k^3\left({1\over e^{\beta_0 k}-1}\right).
\eea

It turns out that the energy density simply scales as $
\mathrm{T}^{4}=\mathrm{T}_0^4
\exp[-4Ht]$,
 as naively expected. This result is reassuring, though, as an unexpected time
behaviour
 of  thermal effects has been observed in another problem \cite{vilenkin} for
the scalar
 field condensates in de Sitter space.

Using the following relation,
\bea
n_k^{b\lambda}(\beta_0)={1\over \beta_0}{\partial\over \partial k}\log Z_k^{b
\lambda}= {1\over \beta_0}{\partial\over \partial k}
\log\biggl (1 - e^{-\beta_0 k}\biggr ),
\eea
 where $ Z_k^{b \lambda}$ is the contribution of the mode $(k,b,\lambda)$ to
the partition function, it results
\bea
\bigl <\mathrm{T_{00}}\bigr >&=&  -{9\over \pi^2}
{e^{-4Ht}\over\beta_0}\int_0^{\infty}\, dk\, k^2
\log\biggl (1-e^{-\beta_0 k}\biggr )\cr
&=& -{9\over \pi^2}\biggl
({e^{-Ht}\over\beta_0}\biggr )^4\int_0^{\infty}\, dx\,  x^2
\log\biggl (1-e^{- x}\biggr ).
\label{<too>}
\eea

{}From eq. (\ref{FCasm}), the free energy at a given time $t$, integrated over
 a comoving patch which had volume $V_0$ at the beginning of inflation,  is
given by
\bea
\mathrm{F(\beta)}&=& -{1\over \beta_0}\log Z(\beta_0)= -{1\over 3}\rho V   =
-{1\over 3}\rho a(t)^3 = -{1\over 3} \bigl <\mathrm{T_{00}}\bigr >V_0e^{3Ht}\cr
 &=& \frac 3 {\pi^2} V_0 T_0^4 e^{-Ht} \int_0^{\infty}  x^2 \log\biggl
(1 - e^{-x}\biggr )dx
\label{FCasm2},\eea
where $\beta=\beta_0e^{Ht}=1/\mathrm{T}$, and $V$ is the volume of the patch at
time
$t$.

\subsection{Computation of the effective potential with an imaginary conformal
time}\label{imaginary}

In flat space, thermal averages of
Green functions can be obtained as vacuum expectation values of the
corresponding operators in a field theory with an imaginary (Euclidian) time,
$\tau \in [0,\beta=1/\mathrm{T}]$, with periodic boundary conditions on the
fields, $A_\mu^a(t=0)=A_\mu^a(t=\beta)$\cite{bernard}.

In de Sitter space, the free
energy, eq. (\ref{FCasm2}), can be recovered from the field theory eq.
(\ref{lag}),
 by analytically continuing the action, in the complex plane of the conformal
time
 $\eta$, eq. (\ref{conformal}).

Let us define $\gamma =\mathrm{Im}\ \eta$. The gauge field equations will be
expressed in terms of $\gamma$, while we will keep the real time $t$
(or equivalently $\mathrm{Re}\
\eta$) to describe the gravitational expansion. All fields are taken as
periodic functions of $\gamma$ with period $\beta_0$.

As in \cite{weiss}, we choose a gauge,
\bea
A^a_\gamma(\gamma,\vec x)= \delta_{a3}\left (\frac {C_0} g + \phi(\vec x)\right
)
\label{jauge}
\eea
where the spatial average of  $\phi(\vec x)$, is $\int d\vec x\phi(\vec x)=0$.
A derivation analogous to the one developped by N. Weiss \cite{weiss}, results
in the partition function:
\bea
Z(\beta)=N\int {\cal D} (g\beta_0 A_\gamma) d A_i^a(\vec x, \gamma)e^{
-\frac 1 2\int_0^{\beta_0}d\gamma \int d^3 x [(\partial_\gamma A_i -\partial_i
A_\gamma +g A_\gamma \times A_i)^2 +B^2]},
\label{Z(beta)}\eea
where
\bea
B^2=\sum_{i<j} G_{ij}^2\,\,.
\eea

The calculation then proceeds exactly as in \cite{weiss} leading to
\bea
\bigl <\mathrm{T_{00}}\bigr >&=& -{3\over \pi^2}\biggl
({e^{-Ht}\over\beta_0}\biggr )^4\int_0^{\infty} dx \biggl \{ x^2 \log\biggl
(1+e^{-2 x} - 2
e^{-x}\cos(\beta_0C_0)\biggr ) \cr
&+& x^2 \log\biggl (1-e^{- x}\biggr )\biggr\}\,,
\label{<toomink>}
\eea
and  to the free energy
\bea
F(\beta)&=&- \frac 1 \beta \log Z(\beta)= -{1\over 3} \bigl
<\mathrm{T_{00}}\bigr >V_0e^{3Ht}\cr
&=&{1\over \pi^2}
V_0 T_0^4 e^{-Ht}\int_0^{\infty} \ dx\ \biggl \{ x^2 \log\biggl (1+e^{-2 x} - 2
e^{-x}\cos(\beta_0C_0)\biggr ) \cr
&+& x^2 \log\biggl (1-e^{- x}\biggr )\biggr \}\cr
&=&{1\over \pi^2}
V T^4 \int_0^{\infty} \ dx\ \biggl \{ x^2 \log\biggl (1+e^{-2 x} - 2
e^{-x}\cos(\beta_0C_0)\biggr ) \cr
&+& x^2 \log\biggl (1-e^{- x}\biggr )\biggr \}\,,
\label{F(beta)}\eea
where the last contribution in eqs. (\ref{<toomink>}) and (\ref{F(beta)})
correspond to the excitations along the $\hat 3$ direction, and where
$V=V_0\exp[3Ht]$.

Notice that taking $C_0=0$ in eq. (\ref{F(beta)}) gives
the relation (\ref{FCasm2}), which was derived in Minkowski space with a
density matrix formalism in real time. This agreement indicates that it is
possible to use
the imaginary time formalism in de Sitter space.

Using the relations:
\bea
\log\biggl (1+e^{-2 x} - 2 e^{-x}\cos(\beta_0 C_0)\biggr ) = 2 \mathrm {Re}
\log\biggl (1-e^{-x} e^{i\beta_0 C_0}\biggr ),
\eea
\bea
\log\biggl (1-e^{-x} e^{i\beta_0 C_0}\biggr )=-\sum_{n=1}^{\infty}{( e^{-x}
e^{i\beta_0 C_0})^n\over n}\,,
\eea
and \cite{tables}
\bea
\sum_{n=1}^{\infty}{\cos nx\over n^4}={\pi^4\over 90 } -{\pi^2 x^2\over 12
}+{\pi  x^3\over 12 } -{ x^4\over 48 }\ ; \,\,\, 0\leq x\leq
2\pi\,\,,\label{serie}
\eea
it results
\bea
\bigl <\mathrm{T_{00}}\bigr >=  {+6\pi^2}\left
({e^{-Ht}\over\beta_0}\right)^4\biggl
\{
{1\over 45}(1+\frac 1 2) -{1\over 24}\biggl\{ 1-\biggl ( \biggl ({\beta_0
C_0\over
\pi}\biggr  )_{\mathrm{Mod} 2} -1\biggr )^2 \biggr \}^2 \biggr\}.
\label{<tooeuc>}
\eea
Finally,  the effective potential of the system is
\bea
\mathrm{V_{eff}(<P(\vec x)>)}=-{1\over 3}\rho={-2\pi^2}\left
({e^{-Ht}\over\beta_0}\right)^4\biggl \{
{1\over 45}(1+\frac 1 2) -{1\over 24}\biggl\{ 1-\biggl ( \biggl({\beta_0
C_0\over
\pi}\biggr )_{\mathrm{Mod} 2} -1\biggr )^2 \biggr \}^2.
\label{<FColi>}
\eea

\noindent The effective potentiel in de Sitter space, eq. (\ref{<FColi>}), is
strikingly similar to the flat space one, eq. (24) in ref. \cite{weiss}. Up to
a global factor $e^{-4Ht}$,
 it does not depend on the real time $t$.
Notice also that the integration over momentum $k$ has been taken from $0$ to
$\infty$ including
the wavelenghts larger than the size of the horizon, $H^{-1}e^{-Ht}$.
 A discussion on this issue will appear in subsection \ref{thermal}.

\subsection{Back to Minkowski space}\label{minkowski}

Is it possible to understand the physical meaning of the $\cos(\beta C_0)$
appearing in eqs. (\ref{<toomink>})-(\ref{<FColi>}). The strange
thermodynamical properties of domains with a non-vanishing $C_0$ have been
analysed in the case of a flat space in \cite{belyaev}.
 One amusing exercise consists in introducing chemical potentials $\mu^+$ and
$\mu^-$ related to the charges $+$ and $-$ respectively,  in equations
eqs. (\ref{densites})-(\ref{FCasm2}). We will proceed briefly along
 these lines just for illustrative purposes,
 as such a choice is void of physical meaning: there is no thermodynamical
stability
unless $\mu^+=\mu^-$ . It leads to the Bose-Einstein distribution functions
\bea
n_k^{(\pm)}(\beta_0)&=&{1\over e^{\beta_0(k- \mu^{\pm})}-1},\cr
n_k^{(3)}(\beta_0)&=&{1\over e^{\beta_0 k}-1}.
\label{densitesC}
\eea
Considering {\it imaginary} chemical potentials, $\mu^+ = -\mu^- =iC_0$,
\bea
n_k^{(+)}+n_k^{(-)}= {1\over \beta_0}{\partial\over \partial k}
\log\biggl (1+e^{-2\beta_0 k} - 2 e^{-\beta_0 k}\cos(\beta_0 C_0)\biggr ),
\eea
 and the  effectif potential for the $+,-$ components writes\footnote{In a
background field,
$A_0^a=i{\delta_{a3} \over g}e^{-Ht}{C_0} $, and with a vanishing chemical
potential,  we would also had obtained eq.(\ref{densitesC})-(\ref{FCasm3}) by
performing {\it formally} the same steps as performed above in the temporal
gauge. However, such a choice  seems to be void of physical meaning. }:
\bea
\mathrm{V_{eff}(<P(\vec x)>)}=\frac 1 {\pi^2} T_0^4 e^{-4Ht} \int_0^{\infty}
x^2
\log\biggl
(1+e^{-2 x} - 2 e^{-x}\cos(\beta_0C_0)\biggr )dx +...\,\,.
\label{FCasm3}\eea
in agreement with (\ref{F(beta)}) using (\ref{signe}).
The meaning of a partition function with an imaginary chemical potential for
the total charge of the system remains to be understood and, on this item, we
have nothing to add to the previous literature \cite{belyaev}-\cite{smilga}.

\section{Instability of the metastale vacua}\label{instability}

\subsection{Absence of thermal decay of metastable vacua}\label{thermal}

The autors of ref. \cite{bhatta} have estimated the interface tension between
the  $Z(N)$
 different vacua in flat space. Their method consisted in combining in one
effective Lagrangian
 the effective potential (at one loop level) $V_{\mathrm{eff}}(<P(\vec
x)>)=F(\beta)/V$ with the kinetic
 energy (tree level). They estimated then the nucleation rate following the
method proposed
 by Coleman \cite{coleman}.
 Using these results, the authors of \cite{korthals} have estimated the
nucleation rate of the true vacuum within a metastable vacuum: the decay rate
per unit space time volume is
 $\Gamma\sim T^4 e^{-S_{\mathrm{eff}}}$, where $S_{\mathrm{eff}} $ turns out to
be
 much larger than $10^3$
in any case.
Their conclusion for flat space was that
 thermal nucleation had no chance to take place.
As previously mentioned, the nucleation rate estimated in ref. \cite{bhatta} is
the subject of
 controversy \cite{belyaev}-\cite{smilga}.
 However, accepting their conclusion for the time being, it is straightforward
to
 extend it to de Sitter space, given the similarity between the effective
action in de Sitter and flat
space.  Indeed, thermal nucleation is even less likely to happen in
inflation. At the very beginning of inflation, $T/H\sim 10^5$, the
nucleation probability in a space time volume, of size $\sim
H^{-4}$, is $\Gamma/H^4\sim (T/H)^4 e^{-S_{\mathrm{eff}}}\ll 1$. The
temperature then falls exponentially and  thermal fluctuations are drastically
damped.
 In short, the relevant space-time volume in de Sitter space
is smaller than in flat space, and the nucleation probability is even
more negligible.

\subsection{Quantum decay of metastable vacua}\label{quantum}

Up to this point we have not considered the quantum fluctuations
related to inflation. We recall that the set of spatial variables
$\vec x$ we use is the set of comoving ones. From the metric, eq.
(\ref{ds}), $|dx/dt|=\exp[-Ht]$ for light, and two points with
coordinates $\vec x$ and $\vec y$ are within one's other horizon only
if $|\vec x-\vec y|<H^{-1} e^{-Ht}$. At a given time $t$, only wave
lengths smaller than $H^{-1} e^{-Ht}$ may be generated by quantum
fluctuations. Longer wave lengths are only fossile remains of quantum
fluctuations that took place earlier. In other words, at a time $t$
the field modes with wave lengths larger than $H^{-1} e^{-Ht}$ can be
considered as classical: they are constant (not fluctuating) in the
whole causally connected domain\footnote{We recall that this
phenomenon is at the origin of the standard understanding of the
classical fluctuations observed in the cosmic background radiation.}.

In eq. (\ref{jauge}), $\phi(\vec x)$ was defined such that $\int d\vec
x\ \phi(\vec x)=0$. However, at a time $t$, and considering a causal
domain around $\vec x_0$, the modes $|\vec k|<He^{Ht}$ have been
``frozen" out by inflation, and the average within this horizon is
$\int_{|\vec x-\vec x_0|<H^{-1}e^{-Ht}} d\vec x\ \phi(\vec x)\ne 0$.
It is convenient to redefine the constant $C_0$ in eq. (\ref{jauge}) in
order to incorporate these ``frozen" modes into a time and spatial
domain dependent $C_0(t,\vec x_0)$. We decompose the gauge field
\bea
A^3_\gamma(\gamma,\vec x)=\frac{C_0(t,\vec x_0)}g +\phi_t^{(\vec
x_0)}(\vec x),
\eea
with
\bea
\int_{|\vec x-\vec x_0|<H^{-1}e^{-Ht}} d\vec x\ \phi_t^{(\vec x_0)}(\vec x)=0,
\eea
where $\gamma$ is again the complexified conformal time and $t$ the
real time.  This implies
\bea
C_0(t,\vec x_0)=C_0 + g\int_{|\vec k|<He^{Ht}} \frac {d\vec
k}{(2\pi)^3}\ \tilde\phi(\vec k)\ e^{i\vec k.\vec x_0},
\eea
where $\tilde\phi(\vec k)$ is defined from
\bea
\phi(\vec x)=\int \frac {d\vec k}{(2\pi)^3} \ \tilde\phi(\vec k)\  e^{i\vec
k.\vec x}.
\label{fourrier}
\eea

We will now consider the change in $C_0(t,\vec x_0)$ during what is
commonly called an e-folding (a time range of lenght $H^{-1}$):
\bea
\delta C_0=C_0(t+H^{-1},\vec x_0)-C_0(t,\vec x_0),
\eea
where from now on, we keep implicit the dependence on $t$ and $\vec
x_0$. It follows that
\bea
\delta C_0=g\int_{He^{Ht}<|\vec k|<He^{Ht+1}} \frac  {d\vec k}{(2\pi)^3}\
\tilde\phi(\vec k)\
  e^{i\vec k.\vec x_0}\,\,.\label{deltac}
\eea

The value of $\tilde\phi(\vec k)$ is undetermined, but its probability
distribution is known. In some sense, inflation is ``measuring" these
modes.

 The quantum average can be computed using the partition function.
{}From eq. (\ref{Z(beta)}) we extract the part that depends on
$\phi(\vec x)$:
\bea
Z_\phi(\beta) = \int {\cal D} \phi \, e^{-\frac 1 2\int_0^{\beta_0}
d\gamma \int d\vec x (\vec \nabla \phi)^2}\label{zphi}.
\eea

In order to compute this integral, we discretize the momentum space in
a finite volume $V$: $$\int \frac {d\vec k}{(2\pi)^3}\to \frac 1 V
\sum_{\vec k}.$$  In this way, the partition function becomes
\bea
Z_\phi(\beta) = \prod_{\vec k} \int d\phi_{\vec k} \, e^{-\frac 1 2
\frac {\beta_0} V |\vec k|^2 |\tilde \phi_{\vec k}^2|},
\eea
and
\bea
\delta C_0 =\frac {g}{V}\sum_{\vec k\in K_t}  \tilde \phi_{\vec k} e^{i\vec
k.\vec x_0}.
\eea
where $K_t$ corresponds to $\vec k$ such that $He^{Ht}<|\vec
k|<He^{Ht+1}$.  The quantum average $<\delta C_0>$ is then
\bea
<\delta C_0>=\frac g {V}\sum_{\vec k\in K_t} e^{i\vec k.\vec x_0}\frac
{\int d \tilde \phi_{\vec k} \,\tilde \phi_{\vec k} e^{-\frac 1 2
\frac {\beta_0} V |\vec k|^2 |\tilde \phi_{\vec k}^2|}}{\int d \tilde
\phi_{\vec k} e^{-\frac 1 2 \frac {\beta_0} V |\vec k|^2 |\tilde
\phi_{\vec k}^2|}} =0\,\,.
\eea
The typical value of $|\delta C_0|$ can be estimated from the squared
average $<\delta C_0^2>$,
\bea
\delta C_0^2=\frac {g^2}{V^2}\sum_{\vec k,\vec k'\,\in K_t }\tilde \phi_{\vec
k} \tilde \phi^*_{\vec k'}e^{i(\vec k - \vec k').\vec x_0}\,\,.
\eea
Given the form of $Z_\phi(\beta)$, only the terms with $\vec k = \vec
k'$ contribute to the average, leading to
\bea
<\delta C_0^2>=\frac {g^2} {V^2}\sum_{\vec k\in K_t}\frac {\int d \tilde
\phi_{\vec k} |\tilde \phi_{\vec k}^2| e^{-\frac 1 2 \frac {\beta_0} V
|\vec k|^2 |\tilde \phi_{\vec k}^2|}}{\int d \tilde \phi_{\vec k}
e^{-\frac 1 2 \frac {\beta_0} V |\vec k|^2 |\tilde \phi_{\vec k}^2|}}
= \frac {g^2} {V^2} \frac V {\beta_0} \sum_{\vec k\in K_t} \frac 1 {|\vec
k|^2}\,\,.
\eea
Taking the continuum limit, we get

\bea
<\delta C_0^2>=T_0\frac {4\pi}{(2\pi)^3} g^2\,\int_{He^{Ht}}^{He^{Ht+1}} dk=
\frac {T_0}g^2 {2 \pi^2} He^{Ht} (e-1)\,\,,
\eea
from which it follows that
\bea
|\delta C_0|\sim \sqrt{T_0 H}e^{\frac {Ht} 2}\sqrt{\frac{e -1}{2
\pi^2}}
\gg (2\pi) T_0\,\,,
\eea
as soon as $He^{Ht}\gg T_0$.

 Since $C_0$ is a periodic variable with a period of $(2\pi) T_0 $, it
results that soon after inflation begins, the value of $C_0(t,\vec
x_0)$ is changed at random after every e-folding by an amount larger
than its period. Consequently, the value of $C_0(t,\vec x_0)$ will
keep no memory of its value one e-folding before and will be
distributed at random in the interval $\in [0,2\pi T_0]$. Finally, the
values $C_0(t,\vec x_0)$ and $C_0(t,\vec x_0 +\delta\vec x_0)$ in the
neighbouring domains are totally uncorrelated since both are
uncorrelated from their common parent domain (the domain where they
were both located one e-folding before).

Further physical insight in the above result can be gained using the
set of variables $\vec x_{phys}=e^{Ht} \vec x$.  In this
parametrization the size of a causal domain remains constant, $\sim
H^{-1}$, while the period of the variable $C(t,\vec x_0)\equiv
C_0(t,\vec x_0)e^{-Ht}$ becomes $2\pi T$, which shrinks to zero as $T$
falls exponentially $T=T_0 e^{-Ht}$. The typical size of quantum
fluctuations overwhelms this period.  In these variables, and defining
$\delta C\equiv \delta C_0 \exp[-Ht]$,
\bea
<\delta C^2>=\frac {T} {2 \pi^2} H (e-1)\gg 4\pi^2 T^2,
\eea
for $T\ll H$. Recall that the present universe comes out from the last
$\simeq 53$ e-foldings during inflation, when the inequality $T\ll H$
is certainly valid.

\subsection{The end of inflation and the reheating.}\label{reheating}
At the end of inflation, the temperature raises
suddenly by the creation of particles to a reheating temperature
$T_{\mathrm{RH}}$.  The effective potential $V_{\mathrm{eff}}(C)$
computed in \cite{weiss} becomes valid with a periodicity  $2\pi
T_{\mathrm{RH}}$.  It is difficult to figure out the evolution of $C$,
a variable defined in thermal equilibrium, during this
out-of-equilibrium reheating. No mechanism is known, however, to
justify the alignement of $C$ on distances larger than the distance of
the causal horizon. In each connected domain, the value of $C$ might
fall into the closest minimum, the latter being uncorrelated with the
minimum in the next causal domain, though.  Then the causal problem
will jump in again, as the CP phases will be randomly distributed in
domains much smaller than the present cluster of galaxies.

\section{Conclusion}

 At the onset of an inflationary era, the putative
temperature-dependent effective potential for the Polyakov line
quickly flattens. The value of the condensate can change and grow,
though, through random quantum fluctuations in de Sitter space, and
can thus be non-zero in a given causal domain when the end of
inflation approaches.

Before reheating, the typical difference in the
value of the condensate between two domains which came out of the
horizon in two consecutive e-foldings is of order $\sim \sqrt{T_0 H
\exp{\frac{H t}2}}$, where $T_0$ was the temperature at the beginning of
inflation.

 Upon reheating, the potential starts to sensibly wiggle again and the value of
the condensate
in a given causal domain might fall into the nearest vaccuum, which may have a
non-trivial
 CP phase  for  large enough gauge groups, such as the standard model
one. The trouble comes from the periodicity of the potential, $\sim 2\pi T_0$.
This is much smaller
than the random quantum change mentioned above, which characterizes for
instance
the $\sim 53$ last e-foldings of inflation which gave birth to the present
universe.
The thermal effects produce a  potential which wraps around itself once in a
period,
while every quantum fluctuation makes the field  wrap around this ``dormant''
 potential many times during one  e-folding.
Two neighbouring domains of our present universe are totally uncorrelated in
what concerns the
value of the condensate, as they are uncorrelated with the common parent domain
produced some
e-foldings before, as soon as the temperature starts to rise again. Our cluster
of galaxies would
result of many domains with opposite CP eigenvalues for the effective field,
which would lead to the random presence of lumps of matter and antimatter,
an untenable result.

\section*{Acknowledgements.}

This work was supported in part by the Human Capital
and Mobility Programme, contract CHRX-CT93-0132, by the CEC Science Project
SC1-CT91-0729 and through funds from CICYT, project AEN93-0673. We acknowledge
Chris Korthals Altes, Jean-Pierre Leroy,
Dominique Levesques, John Madore and Jean Orloff for several inspiring
discussions.

\end{document}